\documentclass[superscriptaddress,showpacs,twocolumn,footinbib,amssymb,aps]{revtex4}

\usepackage{graphicx,dcolumn,bm,amsmath,color,bm}

\begin{document}

\title{Entropic patchiness drives multi-phase coexistence in discotic colloid--depletant mixtures}

\author{\'{A}. Gonz\'{a}lez Garc\'{i}a}
\affiliation{Laboratory of Physical Chemistry, Department of Chemical Engineering and Chemistry, \& Institute for Complex Molecular Systems (ICMS)
Eindhoven University of Technology,P.O. Box $513$, $5600$ MB, Eindhoven, The Netherlands}
\author{H. H. Wensink}
\affiliation{Laboratoire de Physique des Solides - UMR $8502$, Universit\'{e} Paris-Sud and CNRS, $91405$ Orsay Cedex, France}
\author{H. N. W. Lekkerkerker}
\affiliation{Van `t Hoff Laboratory for Physical and Colloid Chemistry, Department of Chemistry \& Debye Institute, Utrecht University, Padualaan $8$, $3584$ CH, The Netherlands}
\author{R. Tuinier}
\affiliation{Laboratory of Physical Chemistry, Department of Chemical Engineering and Chemistry, \& Institute for Complex Molecular Systems (ICMS)
Eindhoven University of Technology,P.O. Box $513$, $5600$ MB, Eindhoven, The Netherlands}
\email{r.tuinier@tue.nl}

\date{\today}



\begin{abstract}
Entropy--driven equilibrium phase behaviour of hard particle dispersions can be understood from excluded volume arguments only. While monodisperse hard spheres only exhibit a fluid--solid phase transition, anisotropic hard particles such as rods, discs, cuboids or boards exhibit various multi--phase equilibria. Ordering of such anisotropic particles increases the free volume entropy by reducing the excluded volume between them. The addition of depletants gives rise to an entropic patchiness represented by orientation--dependent attraction resulting in non--trivial phase behaviour. We show that free volume theory is a simple, generic and tractable framework that enables to incorporate these effects and rationalise various experimental findings. Plate-shaped particles constitute the main building blocks of clays, asphaltenes and chromonic liquid crystals that find widespread use in the food, cosmetics and oil industry. We demonstrate that mixtures of platelets and ideal depletants exhibit a strikingly rich phase behaviour containing several types of three--phase coexistence areas and even a \textit{quadruple} region with four coexisting phases.
\end{abstract}

\maketitle

\section{Introduction}

In the beginning of the 20$^\text{th}$ century Jean Perrin \cite{Perrin1909} showed that colloids behave as large atoms in terms of their statistical thermodynamic properties. As a consequence, this analogy also holds for their phase behaviour \cite{Frenkel2006}: spherical atoms or molecules assume vapour, liquid and solid phase states just as colloidal spheres in a background solvent. The phase behaviour of molecules and colloids depends on the effective interactions between them. The particles typically repel each other at short distances due to excluded volume interaction and may attract one another at larger distances. What makes colloids interesting with respect to molecular systems is that the range and strength of the colloidal interactions as well as colloidal shapes are tuneable \cite{Yethiraj2003}. At the end of the last century it became possible to prepare colloidal dispersions with particles that only interact via a hard--core repulsion  \cite{Pusey1991} For a collection of sufficiently monodisperse colloidal hard spheres the equilibrium phase diagram is a simple one: below a volume fraction of 49\% the dispersion is fluidic and above 55 vol\% it assumes a face--centred cubic crystalline solid phase \cite{Hoover1968,Pusey1986}. This is the most elementary manifestation of an entropy--driven phase transition; it is driven by repulsive (excluded-volume) interactions alone.

Phase transitions in systems of hard colloids with anisotropic shapes typically occur at smaller concentrations. In a seminal paper Onsager \cite{Onsager1942} argued that a suspension of thin hard rods of length $L$ and diameter $D$ undergoes an isotropic--nematic phase transition when the volume fraction of rods is of order $D/L$. Although the rods are orientationally disordered in the isotropic phase and ordered in the nematic phase, the excluded volume interaction between the rods, which scales with $L^2 D$, is smaller in the nematic phase: the available space for a rod increases as the rods align  \cite{Onsager1949}. This is a key thermodynamic property of hard core particles (HCPs): above a certain concentration the entropy increase due to a gain in free volume per particle compensates the orientational entropy loss as the density is not uniform in orientation or position. Free volume entropy also explains the phase transition of long rods from a nematic towards a smectic phase \cite{Frenkel1988,McGrother1996,Bolhuis1997} as well as the liquid crystalline phases encountered in dispersions of platelets \cite{Veerman1992,Wensink2009,Marechal2011} or boards \cite{Camp1997,Vroege2014,Cuetos2017}. In view of their additional orientational degrees of freedom particles with an anisotropic hard core \cite{Dijkstra2016,Meijer2016} exhibit a much richer phase behaviour than their spherical counterparts. Nowadays this phenomenon is usually referred to as \textit{shape entropy} \cite{Glotzer2014}.

\begin{figure*}
\centering
\includegraphics[width=1.8\columnwidth]{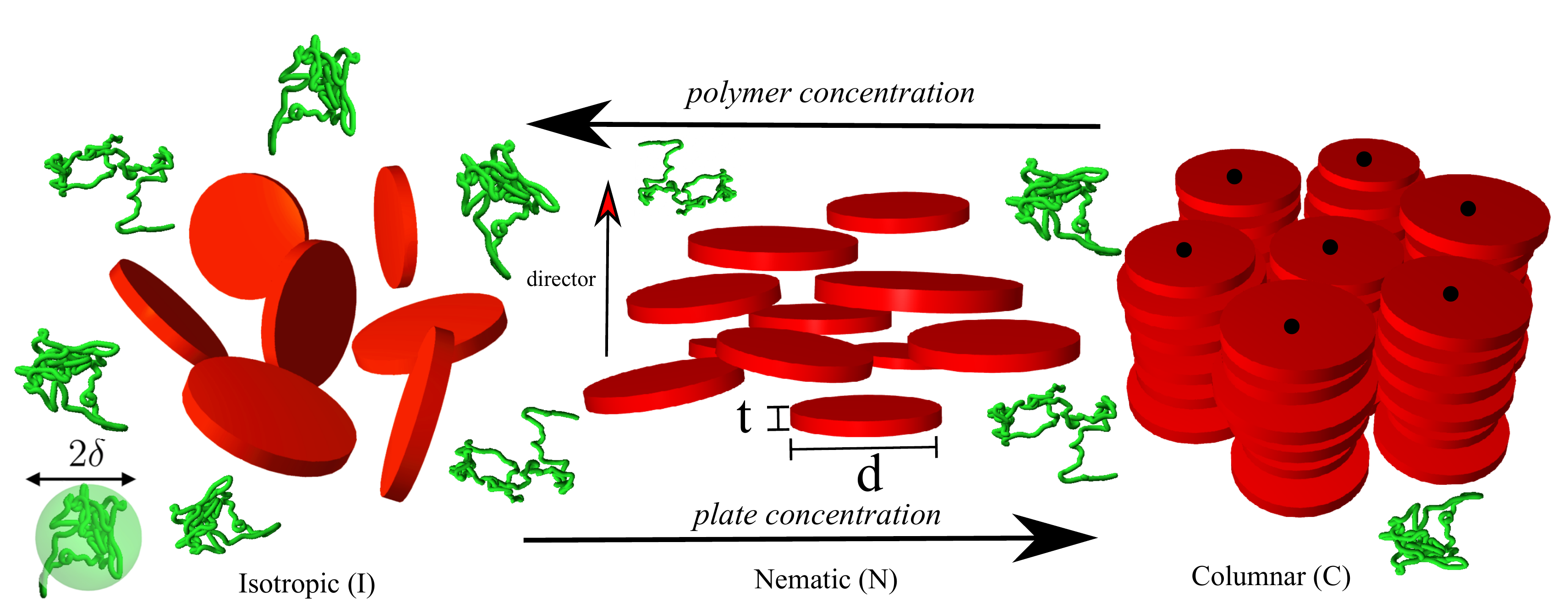}.
\label{fig1}
\caption{Snapshot of an isotropic (I), nematic (N) and columnar (C) phase composed of (red) hard plates with diameter $d$ and thickness $t$ mixed with nonadsorbing polymers (green coils) embedded in a continuum background solvent. In free-volume theory (FVT) the polymer is modeled as a penetrable hard sphere (green transparent sphere) with diameter 2$\delta$.}
\end{figure*}

Adding depletants such as nonadsorbing polymers (or nonadsorbing colloidal particles) to a dispersion of hard--core particles enables a systematic control of the attractions between the HCPs with tunable strength (amount of depletants) and/or range (depletant size) of the attraction, \cite{LekkerkerkerTuinier2011} as was rationalised first by Asakura and Oosawa \cite{Asakura1954}. The depletion effect originates from the excluded volume between the colloidal particles and the depletants \cite{Li-In-On1975,Eisenriegler1983,Gast1983,Sperry1984}. While all direct interactions are repulsive, an effective entropy--induced attraction mediated indirectly via the depletants supplements the direct hard--core repulsion between the colloids. 
Even for simple hard spheres the addition of depletants significantly enriches the phase behaviour of the pure hard sphere dispersion: for small depletants (short--range attraction; depletant radius $\delta \lesssim$
1/3 of the sphere radius $a$) it broadens the fluid--crystal coexistence region and for sufficiently long--ranged attractions, $q = \delta/a \gtrsim 1/3$, a colloidal gas--liquid phase coexistence region appears \cite{Gast1983,Lekkerkerker1992,Meijer1994,Ilett1995}. Besides the isostructural fluid--fluid coexistence an isostructural solid--solid phase transition can be induced by adding tiny hard spherical depletants to an ensemble of bigger hard spheres \cite{Dijkstra1999a,Bolhuis1994}. The addition of depletants imparts a tunable attraction that dramatically modifies the phase behaviour of hard--sphere suspensions.

Intriguingly, in case of anisotropic particles, the depletion attraction becomes inherently orientation--dependent. A better understanding of the consequences of the depletant--mediated entropic  entropic patchiness of hard rod- and plate-shaped colloids on their phase behaviour \cite{Petukhov2017} remains an important issue both from a fundamental scientific standpoint as well in view of many technological applications involving composite discotics. We limit our scope to the most emblematic anisotropic colloidal shapes, \cite{Glotzer2014,Glotzer2017} namely uniaxial rods and platelets, and discuss a selection of experimental and theoretical  observations of liquid crystal phase ordering. For hard rods plus depletants adding nonadsorbing polymers tunes the phase behaviour as follows. For small depletants, $\delta \lesssim D$, the isotropic--nematic (I--N) coexistence region is not affected much, but within the nematic branch there is a possibility of coexistence between two different (isostructural) nematic phases (N--N) \cite{Lekkerkerker1994,Tuinier2007,LekkerkerkerTuinier2011}. At higher rod concentrations a phase with long--ranged uni--dimensional periodic order called a lamellar or smectic (Sm)  phase appears \cite{Frenkel1988,McGrother1996,Bolhuis1997} and there is a nematic--smectic phase coexistence region. At sufficient depletant concentrations this opens up the possibility of three--phase (I--N--Sm) coexistence \cite{Bolhuis1997a}. It may be that N--N coexistence is in practice often superseeded by the appearance of smectic phases \cite{LekkerkerkerTuinier2011}. For somewhat larger depletants ($\delta \sim D$ for $10 \lesssim L/D \lesssim 100$) the addition of nonadsorbing polymers mainly broadens the I--N coexistence region, while N--N coexistence becomes metastable. Experimentally, the broadening of the I--N coexistence upon adding depletants in rodlike dispersions is well known, see for instance \cite{Dogic1997,Edgar2002}. For large depletants the possibility of coexistence between a dilute isotropic and denser isotropic phase (I--I coexistence) appears. In such a case the effective attractions become long--ranged to such a degree that parallel alignment of the rods is not specifically preferred anymore. Theory also predicts several remarkable triphasic I--I--N and I--N--N coexistence regions for certain combinations of $L/D$ and size of the depletants \cite{Tuinier2007}. Three--phase I--I--N coexistence has been observed experimentally in mixtures of rods plus nonadsorbing polymers \cite{Buitenhuis1995,Beck2007} and I--N--N coexistence was detected in mixtures of rodlike virus particle suspensions \cite{Purdy2005}.

The focus in this communication is on developing a simple thermodynamic theory for dispersions of hard platelets plus nonadsorbing polymers, see Fig.~1. Plate-shaped colloids play a key role in certain biological processes (such as the clustering of red blood cells \cite{LekkerkerkerTuinier2011}) as well as constitute a major component of various food and cosmetic products. We explore the phase behaviour of such dispersions and make use of the fact that accurate analytical free energy expressions are available for the fluidic isotropic, nematic and columnar phases of hard discs \cite{Wensink2009}. While a similar theoretical analysis would in principle be possible for rod-depletant mixtures, such an undertaking is severely hampered by the lack of quantitatively reliable information for the thermodynamics of the lamellar phase and we do not further consider the consequences of  entropic patchiness due to rod-shaped objects in the rest of our study. Further motivation to focus on plate-depletant mixtures stems from several important observations of multiphase coexistences reported in experimental studies of platelets plus nonadsorbing polymer chains \cite{Kooij2000,Luan2009} and other types of depletants \cite{Nakato2014,Chen2015,Woolston2015}. We show that many of these phase diagram scenarios can be rationalized using a tractable and generic statistical thermodynamic theory based on simple free-volume concepts \cite{Lekkerkerker1990,Lekkerkerker1992,LekkerkerkerTuinier2011} that we will briefly expound below.

\section{Results and discussion}

Free volume theory (FVT) is used here because it is simple, insightful and accurate. The system of interest contains colloidal platelets, polymer chains and a background solvent in a volume $V$. The platelets are described as hard uniaxial, cylindrical discs having a volume $v_\text{c}= \pi t d^2 /4$, with diameter $d$ and thickness $t$. The geometry of the system of discs and polymer chains can be described using two dimensionless size ratios $\Lambda = d/t$ and $q = 2\delta/ d$. Within FVT the system of interest is held in equilibrium with a reservoir that only contains polymer chains and background solvent. A hypothetical semi--permeable membrane that connects the system and reservoir compartments is permeable for polymer chains and solvent but impermeable for the colloidal platelets. In order to describe the thermodynamic properties we consider the grand potential $\Omega$ quantifying the thermodynamic state of a mixture of hard platelets plus polymers in a background solvent. For such a system an exact thermodynamic expression can be derived, \cite{Aarts2002,Fleer2008,LekkerkerkerTuinier2011} namely ${\Omega}_n = {F}_n - \int_0^{\phi_\text{d}} \alpha_n ( \partial {\Pi}_\text{d}^\text{R} / \partial \phi_\text{d}^{\text{R}'} ) \mathrm{d}\phi_\text{d}^{\text{R}'}$, featuring the Helmholtz free energy $F$, reservoir osmotic pressure ${\Pi}_\text{d}^\text{R}$
and the relative depletant concentration $\phi_d^\text{R}$.  We define the thermal energy $k_B T$ (with $k_B$ Boltzmann's constant) as the unit of energy. The sub--index $n$ refers to the different liquid crystalline phases considered. The quantity $\phi_d^\text{R}$ defines the relative polymer concentration which is implicitly normalised by the coil overlap concentration, whereas the depletant volume  $v_\text{d}=4\pi \delta /3$ is used to render the osmotic pressure dimensionless. Further $\alpha_n$ defines the free volume fraction available for the depletants in the system. It should be noted that $\alpha_n$ is derived specifically for each phase state (isotropic, nematic and columnar) considered, see its definition in the Methods section. 

Since the polymer concentrations needed to induce phase transitions are far below the polymer overlap concentration the polymers are treated as ideal chains which can be described well in terms of penetrable hard spheres \cite{Vrij1976} with respect to depletion--mediated interactions \cite{Eisenriegler1997,Tuinier2000}. Penetrable hard spheres, with a radius $\delta$ that equals the effective depletion thickness, experience a hard interaction with the (hard) colloidal platelets but do not interact with each other, see Fig.~1. This implies ${\Pi}_\text{d}^\text{R}$ can be simplified using Van `t Hoff's law: ${\Pi}_\text{d}^\text{R}=\phi_\text{d}^\text{R}$. A reliable approximation for the free volume fraction can be obtained using scaled particle theory \cite{LekkerkerkerTuinier2011}. Assuming that the free volume corresponds to that of a polymer-free suspension of hard platelets, the dimensionless grand potential is given by
\begin{align}
{\Omega}_n =
	{F}_n - \frac{v_\text{c}}{v_\text{d}}\alpha_n {\Pi}_\text{d}^\text{R}
 \quad \text{.}
\label{eq:Omega}
\end{align}
In order to make progress we require accurate expressions for the Helmholtz energy ${F}_n$ for platelets in the different liquid crystalline phases. We consider isotropic, nematic and columnar phases of hard discs, for which accurate expressions were derived, \cite{Wensink2009} see Methods.

\begin{figure}
\centering
	\includegraphics[width=0.9 \columnwidth]{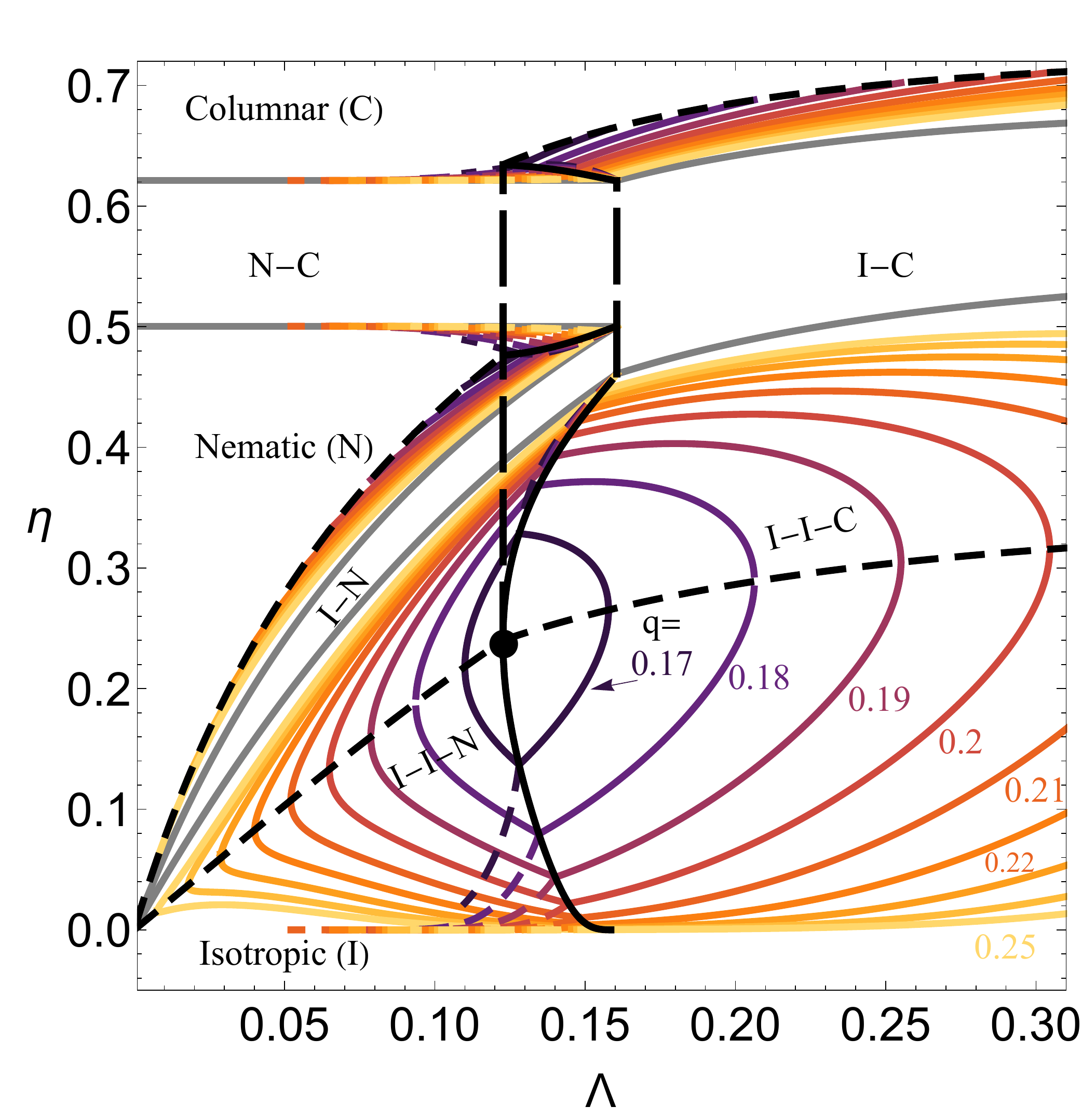}.
	\label{fig2}
	\caption{Multi--phase coexistence $\{\eta,\Lambda\}$ contour plot for indicative relative polymer sizes ($q$) as indicated compared to the phase diagram of a pure platelet suspension (gray solid curves). At $\{\Lambda,q\} \approx \{0.12,0.16\} $ (black dot), a quadruple I--I--N--C curve arises (black solid curve), which spans from $\Lambda \approx 0.12$ (long--dashed, black, vertical line) to $\Lambda \approx 0.16$ (long--dashed, black, vertical line). Black dashed curves correspond to the critical--end--points for I--I--N coexistence ($\Lambda  \lesssim 0.12$) and to the critical--end--points for the I--I--C coexistence ($\Lambda  \gtrsim 0.12$).}
\end{figure}

\begin{figure*}\centering
	\includegraphics[width=1.8\columnwidth]{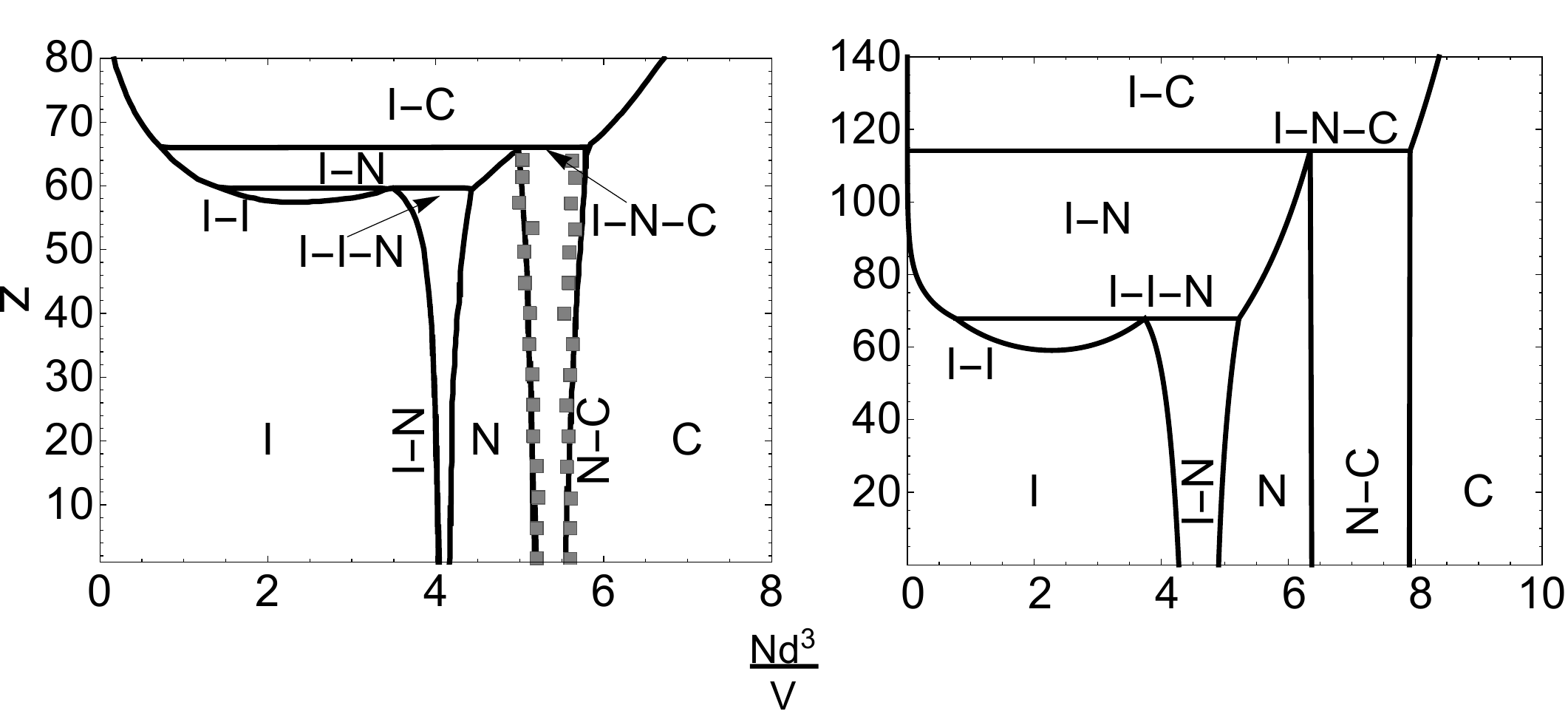}.
	\label{fig3}
	\caption{Comparison of computed phase diagram for a mixtures of colloidal platelets plus nonadsorbing polymers (simplified as mutually penetrable hard spheres) for $\Lambda$=0.1 and $q$=0.2. Left: hybrid approach of free volume theory with Monte Carlo computer simulation results for the equations of states of the various phase states (curves and data represent two methods) for cut spheres plus depletants \cite{Zhang2002}. Right: our free volume theory calculation results for discs plus depletants.}
\end{figure*}

The phase behaviour of pure platelets can be computed from standard thermodynamics (equality of chemical potential and osmotic pressure in each of the coexisting phases; see Methods) from the free energy expressions (see Methods) of Eqs.~\eqref{eq:FreeEnIsoPlate}, \eqref{Fnem} and \eqref{eq:FCol} and it has been demonstrated \cite{Wensink2009} that these are close to Monte Carlo computer simulation results \cite{Wensink2009}. The isotropic--nematic (I--N), isotropic--columnar (I--C) and nematic--columnar (N--C) coexistences are plotted in Fig.~2 as the light grey curves (marked by the texts I--N, N--C and I--C in between coexisting curves); coloured curves will be discussed later on. Starting from the limit of infinitely thin plates ($\Lambda \longrightarrow 0$) the volume fractions $\eta$ for a suspension of discs at N--C coexistence hardly varies with increasing $\Lambda$. In contrast the I--N coexistence volume fractions strongly increase with increasing $\Lambda$ until they merge with the N--C coexistence at $\Lambda \approx 0.16$, beyond which the nematic region ceases to exist and only I--C coexistence survives for $\Lambda   \gtrsim  0.16$.

Next, we study the influence of added depletants on the phase behaviour using Eq.~(1), from which the chemical potential of the discs and osmotic pressure of the mixture can be computed at given depletant reservoir concentration $\phi_\text{d}^\text{R}$. Earlier computer simulations \cite{Bates2000}, combined computer simulation plus free volume theory approaches \cite{Bates2000,Zhang2002,Zhang2002a} and fundamental measure theory \cite{Aliabadi2016} have hinted at the richness of the phase behaviour. In Fig.~3 (left panel) we have redrawn the hybrid approach of Monte Carlo computer simulation and theory results by Zhang, Reynolds and van Duijneveldt \cite{Zhang2002} in terms of the fugacity $z$ (=$6\phi_\text{d}^\text{R}$/[$\pi q^3$]) versus platelet concentration $Nd^3/V=4\eta/(\pi \Lambda)$, with $N$ the number of platelets in the system of volume $V$. At zero depletant concentration ($z=0$) the I--N and N--C coexistences of pure hard platelets are recovered. Upon increasing the depletant concentration (higher $z$ values) the I--N coexistence slowly widens until an I--I--N triple line appears. At relatively small platelet concentrations there is coexistence between two isotropic fluid phases (I--I). Such an isostructural coexistence region has been reported recently by Chen \textit{et al.} \cite{Chen2015} for aqueous suspensions of Zirconium phosphate platelets with added (very dilute) silica spheres as depletants. At higher $z$ values the N--C coexistence also widens a bit until it hits an I--N--C triple line. The hybrid computer simulation--theory results are (Fig. 3; left panel) qualitatively reproduced by our theoretical free volume calculations (Fig.~3; right panel). The correspondence of Fig.~3 benchmarks our FVT approach. In the S.I. a comparison is also shown at another relative depletant size $q$.

\begin{figure*}
	\includegraphics[width= 1.8 \columnwidth]{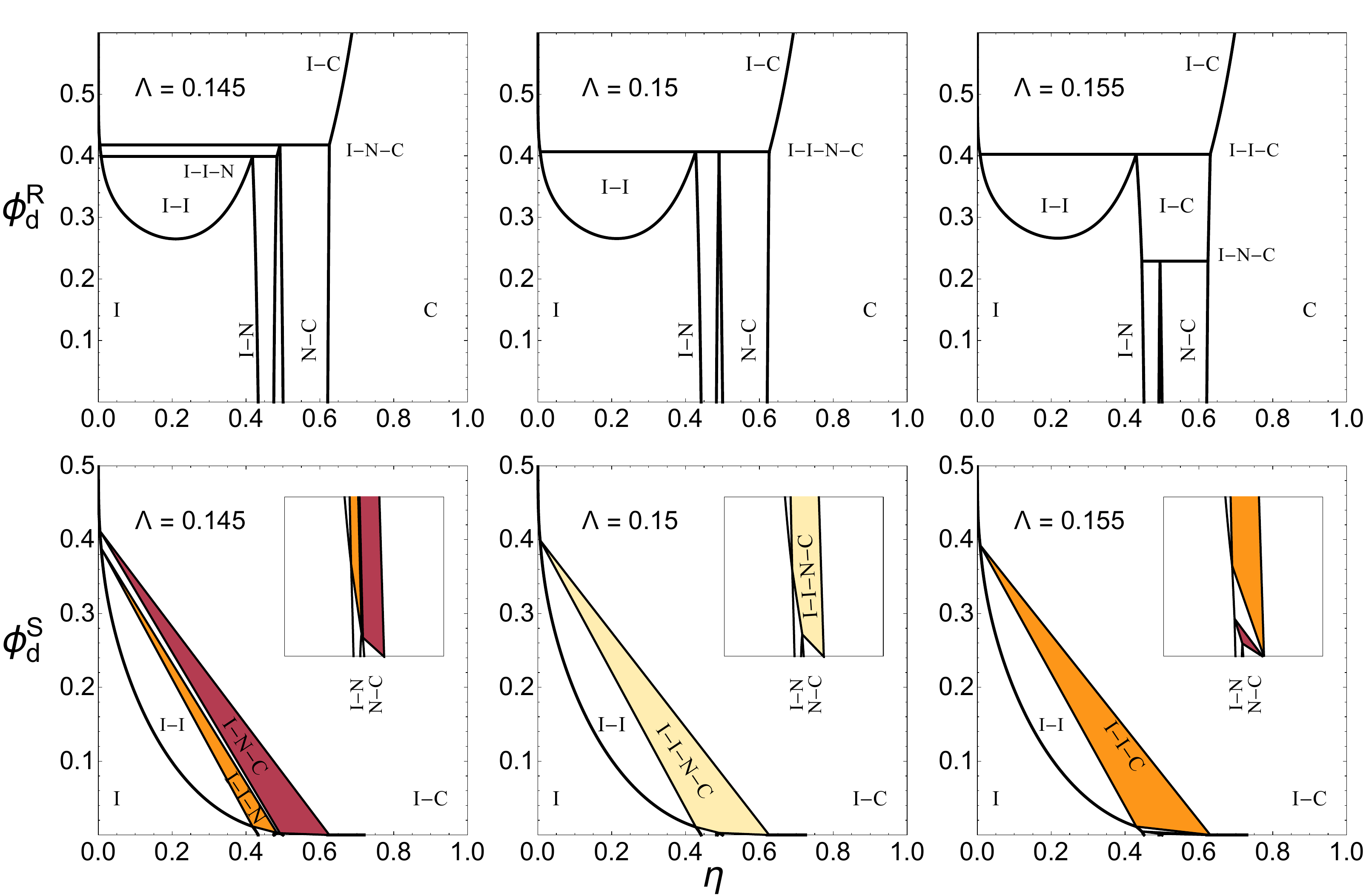}.
	\label{fig4}
	\caption{Collection of phase diagrams in the $\{\eta,\phi_\text{d}^\text{R}\}$ (top panels; reservoir representation) and $\{\eta,\phi_\text{d}^\text{S}\}$ (lower panels) phase spaces for hard platelets plus added depletants at $q$=0.215 as a function of aspect ratio $\Lambda$. Horizontal lines (top panels) indicate multiple--phase coexistences. Inset plots zoom into the low--depletant concentration regime.}
\end{figure*}

A major advantage of FVT is that it enables a facile and systematic exploration of the various possible phase states and coexistences by variation of the relative depletant size ($q$) and relative platelet thickness ($\Lambda$). An example is given in Fig.~4 for $q$ =0.215 upon varying $\Lambda$ from 0.145 to 0.155, both in terms of the reservoir depletant concentration (upper panels) along the ordinate (as in Fig.~3) as well as for the system depletant concentrations (lower panels). The depletant concentration in the system follows from the reservoir concentration as $\phi_d^\text{S}=\alpha_n\phi_\text{d}^\text{R}$. At $\Lambda$=0.145 (upper left panel) we find a similar situation as in Fig.~3 but with the two triple lines (I--I--N and I--N--C) close to each other. In the lower panel it is shown that these triple lines become triple \textit{regions} (coloured orange and red, respectively) when the depletant concentration in the system is considered, as was similarly demonstrated for hard spheres or rods plus depletants \cite{Lekkerkerker1992,Lekkerkerker1994}. The transition from a triple line to a triple region originates from the partitioning of the depletants over the coexisting phases, a key strong element of FVT. For $\Lambda$=0.155 (upper right panel) the two triple lines have crossed; the I--I--N triple line now has become an I--I--C triple line and the I--N--C line has dropped to lower depletant concentration $\phi_\text{d}^\text{R}$. As a result, at the transition point at $\Lambda=$0.150 (middle upper panel) a \textit{quadruple} I--I--N--C coexistence appears. In the system representations (lower panels) the four--phase coexistence line becomes a quadruple \textit{region}. Such four--phase coexistence was reported experimentally for a mixture of sterically stabilised gibbsite platelets mixed with polydimethylsiloxane (PDMS) polymer chains in toluene, \cite{Kooij2000} but was not yet predicted by theory or simulation. The detected four--phase region was assigned by the authors to polydispersity effects. Using FVT we can rationalise such a four--phase coexistence in a platelet plus depletant mixture without the need to take into account polydispersity effects or sedimentation \cite{Wensink2004}. In a different system of aqueous dispersions of rather polydisperse positively charged Mg$_2$Al layered double hydroxide platelets  mixed with polyvinyl pyrrolidone polymers, Luan \textit{et al.} \cite{Luan2009} also reported four--phase coexistence. These experimental examples corroborate the possibility of the quadruple region reported in this study.

We return to Fig.~2 where, besides the phase coexistence regions of hard platelets (gray, solid curves), also contour plots have been added for various depletant sizes, expressed via $q$. Each solid coloured curve corresponds to a triple coexistence (I--I--N or I--I--C), while dashed curves correspond to the two possible I--N--C (black dashed)  triple coexistences (above or below the I--I containing triple point, see Fig.~3). Hence we find a plethora   of different triple lines. When the two solid curves collapse, an I--I--N--C quadruple line is found (solid black curve). Hence, all solid coloured curves to the left of the quadruple curve correspond to I--I--C coexistences, whereas the ones on the right correspond to I--I--N triple lines. As expected, the I--N--C triple lines also collapses to the quadruple line. All triple point curves run from the quadruple line to the depletant--free limit with increasing $q$, and indicates the dramatic effect of depletion--mediated attraction between the discs. The solid, black, dashed curves are the triple I--I--N and I--I--C end point regions: they mark the points at which the I--I critical point matches with the I--I--N or I--I--C triple point. Hence, the black curves correspond to the points where the two I--I--N or I--I--C triple point branches collapse. The location at which the critical I--I--N end point curves match with the critical I--I--C end points defines the I--I part of the critical end point of the quadruple (big black dot), which acts here as reference point for all other curves. Finally, the two long--dashed vertical curves correspond to the triple point I--N--C for platelets in the absence of depletants and to the quadruple critical end point for colloidal platelet--polymer mixtures. This indicates that the isotropic--columnar coexistence occurs at lower $\Lambda$ due to the addition of depletants.

Fundamental insight into multi-phase coexistence is of considerable scientific relevance in view of a range of applications where fluid stability and homogeneity needs to be guaranteed or it can be explored to optimize particle size fractionation and purify colloidal composite materials in an efficient, sustainable and low-cost manner, \cite{Park2010} thereby circumventing energy consuming separation technology. Regions where multiple phases coexist are of relevance for developing switchable or multistable materials \cite{Cox2001,Ahart2008,Park2013} whose properties change upon applying an external mechanical or electro-magnetic field. Last not least, a detailed understanding of the phase behaviour of colloid--polymer mixtures is important for the development of photovoltaic cells \cite{Saunders2008} and polymer--based solar cells \cite{Kouijzer2013}.

In this communication we have shown that shape entropy is an emergent property that leads to a rich phase behaviour of mixtures of hard discs plus nonadsorbing polymers including a wide range of triple coexistences and a quadruple region, hitherto unexplored by theory or simulation. The findings derived from free volume theory (FVT) for discs plus depletants are supported by experiments and are in line with numerical Monte Carlo computer simulation--theory data. 

Contrary to computer simulation or extensive density functional treatments \cite{Wu2007}, FVT offers an efficient and versatile tool to systematically explore parameter phase space, spanned by the colloid aspect ratio and colloid-depletant size ratio, and scrutinize various  phase behaviour scenarios that may subsequently serve as useful benchmarks for more in-depth experimental and simulation explorations. FVT reveals those phase space regions where specific phase coexistences may be found in experimental work or computer simulation studies. 

We have demonstrated that free-volume theory (FVT) enables a simple, tractable and insightful analysis of the phase behaviour of colloid--polymer mixtures while maintaining accurate predictions on the semi--quantitative level with only two size ratios as system input parameters. Since FVT captures the essential physics of depletion-mediated entropic patchiness of simple rod- or disk-shaped building blocks, the predictions are expected to be qualitatively fully equivalent to those emerging from more accurate but technically more demanding treatments. Future efforts will be directed towards gaining a deeper understanding of isostructural I--N--N triple equilibria that have been reported experimentally in mixtures of platelets plus depletants \cite{Luan2009,Nakato2014}. Other types of triple and quadruple phase-coexistences involving several fluid and liquid crystalline phases with distinctly different symmetries will be investigated as well.

\section{Methods}
\subsection*{Free energy expressions of hard discs}
For the  isotropic state the volume fraction ($\eta$) dependence of the (dimensionless) excess free energy of an isotropic ensemble of cylindrical hard discs can be written as \cite{Wensink2009}:
\begin{align}
{F}_\text{I} = 
\frac{2}{\pi}\frac{\eta^2}{\Lambda}\,G\,\Theta_\text{I}(\Lambda) \quad \text{,}
\label{eq:FreeEnIsoPlate}
\end{align}
where $G = (4-3 \eta)/(4 [1-\eta]^2)$ is a virial renormalization approximation proposed by Parsons \cite{Parsons1979} and Lee \cite{Lee1987} for virial coefficients beyond the pair level. {We omit the ideal contribution, $\text{cnst}\times  \eta + \eta \ln\eta- \eta$, in all free energies.} The normalised ensemble averaged excluded volume in the isotropic phase is given by:
\begin{align*}
\Theta_\text{I}  & = \frac{\pi ^2}{8} + \left(\frac{3 \pi }{4}+\frac{\pi ^2}{4}\right) \Lambda + \frac{\pi  \Lambda ^2}{2}  \quad \text{.}
\end{align*}
In the nematic state one should account for the alignment of the hard platelets. The orientational distribution function (ODF), describing the orientational probability of the plate normals with respect to the nematic director is assumed to be a simple Gaussian, \cite{Odijk1986,Vroege1992} $\sim ({\kappa}/{4\pi})\exp\left[-\kappa\theta^2 /2\right] $. The ODF only depends on the polar angle $\theta$ between two rods \cite{LekkerkerkerTuinier2011}. Hence one arrives at:
\begin{align}
\frac{{F}_\text{N}}{\eta} = 
\sigma_\text{N} +
\frac{2}{\pi}\frac{\eta}{\Lambda}G\Theta_\text{N} \quad \text{,}
\label{Fnem}
\end{align}
with $\kappa =\pi \eta^2 G^2/(4\Lambda)$, a variational parameter indicating the degree of nematic alignment as follows from minimizing the free energy with respect to the ODF. The free energy Eq.~\ref{Fnem} now includes a rotational entropy term $\sigma_\text{N} =  \ln \kappa-1$ and a normalised excluded volume:
\begin{align*}
\Theta_\text{N} &= \frac{1}{2}\pi ^{3/2}{\kappa}^{-1/2}+2 \pi  \Lambda \quad \text{.}
\end{align*}

For the columnar phase, cell theory \cite{Lennard1937} pre-supposes the disc columns to accommodate a single--occupancy bidimensional hexagonal lattice while the discs experience strictly uni-dimensional intracolumnar fluid order with asymptotically weak orientational freedom. A major advantage of this approach is that it yields a closed expression for the free energy \cite{Wensink2009}:
\begin{align}
\begin{split}
\frac{{F}_\text{C}}{\eta} &= 
-2 \ln \left(1-\frac{1}{\text{$\Delta $C}}\right) \\
&+2 \ln \left(\frac{3 \text{$\Delta $C}^2 \phi^*}{2 \Lambda  \left(1-\text{$\Delta $C}^2 \phi^*\right)}\right) \\
&-\ln \left(\frac{1}{3} -\frac{\text{$\Delta $C}^2 \phi^*}{3}\right)-2 \quad \text{,}
\end{split}
\label{eq:FCol}
\end{align}
with $\Delta C = [\sqrt[3]{2} K^{2/3}-\sqrt[3]{3} 4 \phi^*]/[6^{2/3} \sqrt[3]{K} \phi^*]$
and $K = \sqrt{3 {\phi^*}^3 (243 \phi^*+32)}\, +\,27 {\phi^*}^2$, where $\phi^*= {\eta}/{\eta_\text{cp}}$, with $\eta_\text{cp}$ the closest packing fraction of columnar discs \cite{Malijevski} $\eta_\text{cp}=\pi\sqrt{3}/6 \approx 0.907$.

\subsection{Free volume fraction for depletants in a suspension of hard platelets}
In order to compute the grand potential from Eq.~(1) we need the free volume fraction of depletants in the pure platelet suspension. This can be derived from scaled particle theory \cite{LekkerkerkerTuinier2011} and yields: 
\begin{align}
\alpha_n &= (1-\eta )	\exp \left[-Q\right]\exp \left[-\frac{v_\text{d}}{v_\text{c}}{\Pi}_n^\text{o} \right] ,
\label{eq:alfa}
\end{align}
in which $v_\text{d}$ is the depletant volume. The quantity ${\Pi}_n^\text{o}$ can be computed from the free energy $F_n$ given in the previous subsection. The quantity $Q$ corresponds to a shape-dependent immersion free energy term for inserting depletants into the colloidal dispersion and reads:
\begin{align*}
Q &= q \left(\frac{1}{\Lambda }+\frac{\pi  q}{2 \Lambda }+q+2\right) y(\eta) \nonumber \\ 
&+ 2 q^2 \left(\frac{1}{4 \Lambda ^2} +\frac{1}{\Lambda }+1\right) [y(\eta)]^2  \text{,}
\end{align*}
with
\begin{align*}
y(\eta) = \frac{\eta}{1-\eta} \quad .
\end{align*}

\subsection{Phase coexistence calculations}
\label{phasecoexcalc}
Coexistence between different phases is established if the chemical potentials of the platelets $\mu$ and osmotic pressures of the platelet--depletant mixtures $\Pi$ are equal in those phases:
\begin{align}
{\Pi}_{i} = {\Pi}_{j}\quad
\text{, and}
\quad
{\mu}_{i} = {\mu}_{j}  \quad \text{,}
\end{align}
where the subscripts $\{i,j\}$ denote the possible states of the system: isotropic (I), nematic (N), or columnar (C).

The chemical potential and osmotic pressure can be calculated from the grand potential in Eq.~(1) using 
\begin{align}
{\mu}_n=\left(\frac{\partial{\Omega}_n}{\partial\eta}\right)_{T,V,N_{\text{d}}^{\text{R}}} \quad\text{and}\quad {\Pi}_n = \eta{\mu}_n - {\Omega}_n ,
\end{align}
where $N_{\text{d}}^{\text{R}}$ is the number of depletants in the reservoir of volume $V$. 

An isostructural phase coexistence of two phases with the same symmetry is characterized by a critical point, which can be calculated from the conditions:
\begin{align}
\left(\frac{\partial{\mu}_n}{\partial\eta}\right)_{T,V,N_d^R} = \left(\frac{\partial^2{\mu}_n}{\partial\eta^2}\right)_{T,V,N_d^R} = 0 \quad \text{.}
\end{align}
Three phases coexist when the following condition is met:
\begin{align}
{\Pi}_{i} = {\Pi}_{j} = {\Pi}_{m} \quad
\text{, and} \quad
{\mu}_{i} = {\mu}_{j} = {\mu}_{m}  \quad ,
\end{align}
where $m$ is a certain phase identity (I, N or C). Likewise, a four-phase coexistence naturally follows from the condition:
\begin{align}
{\Pi}_{i} = {\Pi}_{j} = {\Pi}_{m} = {\Pi}_{n}\quad
\text{, and} \quad
{\mu}_{i} = {\mu}_{j} = {\mu}_{m} = {\mu}_{n}  \quad ,
\end{align}
where $n$ is a certain phase state, where two of the four coexisting phases have the same phase identity.

\subsection*{Data availability}
\label{data}
The data that support the findings of this work are available from the corresponding authors upon request. We can also provide Wolfram Mathematica files that were prepared to calculate the phase diagrams.

\bibliographystyle{apsrev}

\begin{thebibliography}{10}
	\expandafter\ifx\csname url\endcsname\relax
	\def\url#1{\texttt{#1}}\fi
	\expandafter\ifx\csname urlprefix\endcsname\relax\def\urlprefix{URL }\fi
	\providecommand{\bibinfo}[2]{#2}
	\providecommand{\eprint}[2][]{\url{#2}}
	
	\bibitem{Perrin1909}
	\bibinfo{author}{Perrin, J.}
	\newblock \bibinfo{title}{Mouvement brownien et r\'{e}alit\'{e} mol\'{e}culaire
		[brownian movement and molecular reality]}.
	\newblock \emph{\bibinfo{journal}{Ann. Chim. Phys. 8i\`{e}me s\'{e}rie}}
	\textbf{\bibinfo{volume}{18}}, \bibinfo{pages}{5--114}
	(\bibinfo{year}{1909}).
	
	\bibitem{Frenkel2006}
	\bibinfo{author}{Frenkel, D.}
	\newblock \bibinfo{title}{Colloidal encounters: a matter of attraction}.
	\newblock \emph{\bibinfo{journal}{Nature}} \textbf{\bibinfo{volume}{314}},
	\bibinfo{pages}{768} (\bibinfo{year}{2006}).
	
	\bibitem{Yethiraj2003}
	\bibinfo{author}{Yethiraj, A.} \& \bibinfo{author}{van Blaaderen, A.}
	\newblock \bibinfo{title}{A colloidal model system with an interaction tunable
		from hard sphere to soft and dipolar}.
	\newblock \emph{\bibinfo{journal}{Nature}} \textbf{\bibinfo{volume}{421}},
	\bibinfo{pages}{513--517} (\bibinfo{year}{2003}).
	
	\bibitem{Pusey1991}
	\bibinfo{author}{Pusey, P.~N.}, \bibinfo{author}{Van~Megen, W.},
	\bibinfo{author}{Underwood, S.~M.}, \bibinfo{author}{Bartlett, P.} \&
	\bibinfo{author}{Ottewill, R.~H.}
	\newblock \bibinfo{title}{Colloidal fluids, crystals and glasses}.
	\newblock \emph{\bibinfo{journal}{Physica A}} \textbf{\bibinfo{volume}{176}},
	\bibinfo{pages}{16--27} (\bibinfo{year}{1991}).
	
	\bibitem{Hoover1968}
	\bibinfo{author}{Hoover, W.~G.} \& \bibinfo{author}{Ree, F.~H.}
	\newblock \bibinfo{title}{Melting transition and communal entropy for hard
		spheres}.
	\newblock \emph{\bibinfo{journal}{J. Chem. Phys.}}
	\textbf{\bibinfo{volume}{49}}, \bibinfo{pages}{3609} (\bibinfo{year}{1968}).
	
	\bibitem{Pusey1986}
	\bibinfo{author}{Pusey, P.~N.} \& \bibinfo{author}{Van~Megen, W.}
	\newblock \bibinfo{title}{Phase behaviour of concentrated suspensions of nearly
		hard colloidal spheres}.
	\newblock \emph{\bibinfo{journal}{Nature}} \textbf{\bibinfo{volume}{320}},
	\bibinfo{pages}{340} (\bibinfo{year}{1986}).
	
	\bibitem{Onsager1942}
	\bibinfo{author}{Onsager, L.}
	\newblock \bibinfo{title}{Anisotropic solutions of colloids}.
	\newblock \emph{\bibinfo{journal}{Phys. Rev.}} \textbf{\bibinfo{volume}{62}},
	\bibinfo{pages}{558} (\bibinfo{year}{1942}).
	
	\bibitem{Onsager1949}
	\bibinfo{author}{Onsager, L.}
	\newblock \bibinfo{title}{The effects of shape on the interaction of colloidal
		particles}.
	\newblock \emph{\bibinfo{journal}{Ann. N.Y. Acad. Sci.}}
	\textbf{\bibinfo{volume}{51}}, \bibinfo{pages}{627} (\bibinfo{year}{1949}).
	
	\bibitem{Frenkel1988}
	\bibinfo{author}{Frenkel, D.}, \bibinfo{author}{Lekkerkerker, H. N.~W.} \&
	\bibinfo{author}{Stroobants, A.}
	\newblock \bibinfo{title}{Thermodynamic stability of a smectic phase in a
		system of hard rods}.
	\newblock \emph{\bibinfo{journal}{Nature}} \textbf{\bibinfo{volume}{332}},
	\bibinfo{pages}{822} (\bibinfo{year}{1988}).
	
	\bibitem{McGrother1996}
	\bibinfo{author}{McGrother, S.~C.}, \bibinfo{author}{Williamson, D. C.} \& \bibinfo{author}{Jackson, G.}
	\newblock \bibinfo{title}{A re--examination of the phase diagram of hard spherocylinders}.
	\newblock \emph{\bibinfo{journal}{J. Chem. Phys.}}
	\textbf{\bibinfo{volume}{104}}, \bibinfo{pages}{6755-6771} (\bibinfo{year}{1996}).
	
	\bibitem{Bolhuis1997}
	\bibinfo{author}{Bolhuis, P.~G.} \& \bibinfo{author}{Frenkel, D.}
	\newblock \bibinfo{title}{Tracing the phase boundaries of hard
		spherocylinders}.
	\newblock \emph{\bibinfo{journal}{J. Chem. Phys.}}
	\textbf{\bibinfo{volume}{106}}, \bibinfo{pages}{666} (\bibinfo{year}{1997}).
	
	\bibitem{Veerman1992}
	\bibinfo{author}{Veerman, J. A.~C. } \& \bibinfo{author}{Frenkel, D.}
	\newblock \bibinfo{title}{Phase behavior of disklike hard--core mesogens}.
	\newblock \emph{\bibinfo{journal}{Phys.  Rev.  A}} \textbf{\bibinfo{volume}{45}},
	\bibinfo{pages}{5632-5648} (\bibinfo{year}{1992}).

	\bibitem{Wensink2009}
	\bibinfo{author}{Wensink, H.~H.} \& \bibinfo{author}{Lekkerkerker, H. N.~W.}
	\newblock \bibinfo{title}{Phase diagram of hard colloidal platelets: a
		theoretical account}.
	\newblock \emph{\bibinfo{journal}{Mol. Phys.}} \textbf{\bibinfo{volume}{107}},
	\bibinfo{pages}{2111} (\bibinfo{year}{2009}).

	\bibitem{Marechal2011}
\bibinfo{author}{Marechal, M.}, \bibinfo{author}{Cuetos, A.}, \bibinfo{author}{Mart\'{i}nez-Haya, B.} \& \bibinfo{author}{Dijkstra, M.}
\newblock \bibinfo{title}{Phase behavior of hard colloidal platelets using free energy calculations}.
\newblock \emph{\bibinfo{journal}{J. Chem. Phys.}}
\textbf{\bibinfo{volume}{134}}, \bibinfo{pages}{094501} (\bibinfo{year}{2011}).

	\bibitem{Camp1997}
\bibinfo{author}{Camp, P.J.}  \& \bibinfo{author}{Allen, M.P.}
\newblock \bibinfo{title}{Phase diagram of the hard biaxial ellipsoid fluid}.
\newblock \emph{\bibinfo{journal}{J. Chem. Phys.}}
\textbf{\bibinfo{volume}{106}}, \bibinfo{pages}{6681-6688} (\bibinfo{year}{1997}).


	\bibitem{Vroege2014}
\bibinfo{author}{Vroege, G.J.}
\newblock \bibinfo{title}{Biaxial phases in mineral liquid crystals}.
\newblock \emph{\bibinfo{journal}{Liquid Cryst.}} \textbf{\bibinfo{volume}{41}},
\bibinfo{pages}{342-352} (\bibinfo{year}{2014}).

	
	\bibitem{Cuetos2017}
	\bibinfo{author}{Cuetos, A.}, \bibinfo{author}{Dennison, M.},
	\bibinfo{author}{Masters, A.} \& \bibinfo{author}{Patti, A.}
	\newblock \bibinfo{title}{Phase behaviour of hard board-like particles}.
	\newblock \emph{\bibinfo{journal}{Soft Matter}} \textbf{\bibinfo{volume}{13}},
	\bibinfo{pages}{4720} (\bibinfo{year}{2017}).
	
	\bibitem{Dijkstra2016}
	\bibinfo{author}{Dussi, S.} \& \bibinfo{author}{Dijkstra, M.}
	\newblock \bibinfo{title}{Entropy-driven formation of chiral nematic phases by
		computer simulations}.
	\newblock \emph{\bibinfo{journal}{Nat. Commun.}} \textbf{\bibinfo{volume}{7}},
	\bibinfo{pages}{11175} (\bibinfo{year}{2016}).
	
	\bibitem{Meijer2016}
	\bibinfo{author}{Meijer, J.~M.} \emph{et~al.}
	\newblock \bibinfo{title}{Observation of solid-solid transitions in 3d crystals
		of colloidal superballs}.
	\newblock \emph{\bibinfo{journal}{Nat. Commun.}} \textbf{\bibinfo{volume}{8}},
	\bibinfo{pages}{14352} (\bibinfo{year}{2017}).
	
	\bibitem{Glotzer2014}
	\bibinfo{author}{van Anders, G.}, \bibinfo{author}{Klotsa, D.},
	\bibinfo{author}{Ahmed, N.~K.}, \bibinfo{author}{Engel, M.} \&
	\bibinfo{author}{Glotzer, S.}
	\newblock \bibinfo{title}{Understanding shape entropy through local dense
		packing}.
	\newblock \emph{\bibinfo{journal}{PNAS}} \textbf{\bibinfo{volume}{111}},
	\bibinfo{pages}{E4812--E4821} (\bibinfo{year}{2014}).
	
	\bibitem{LekkerkerkerTuinier2011}
	\bibinfo{author}{Lekkerkerker, H. N.~W.} \& \bibinfo{author}{Tuinier, R.}
	\newblock \emph{\bibinfo{title}{Colloids and the Depletion Interaction}}
	(\bibinfo{publisher}{Springer, Dordrecht}, \bibinfo{year}{2011}).
	
	\bibitem{Asakura1954}
	\bibinfo{author}{Asakura, S.} \& \bibinfo{author}{Oosawa, F.}
	\newblock \bibinfo{title}{On interaction between two bodies immersed in a
		solution of macromolecules}.
	\newblock \emph{\bibinfo{journal}{J. Chem. Phys.}}
	\textbf{\bibinfo{volume}{22}}, \bibinfo{pages}{1255--1256}
	(\bibinfo{year}{1954}).
	
	\bibitem{Li-In-On1975}
	\bibinfo{author}{Li-In-On, R.}, \bibinfo{author}{Vincent, B.} \&
	\bibinfo{author}{Waite, F.~A.}
	\newblock \bibinfo{title}{Stability of sterically stabilized dispersions at
		high polymer concentrations}.
	\newblock \emph{\bibinfo{journal}{ACS Symp. Ser.}}
	\textbf{\bibinfo{volume}{9}}, \bibinfo{pages}{165} (\bibinfo{year}{1975}).
	
	\bibitem{Eisenriegler1983}
	\bibinfo{author}{Eisenriegler, E.}
	\newblock \bibinfo{title}{Dilute and semidilute polymer solutions near an
		adsorbing wall}.
	\newblock \emph{\bibinfo{journal}{J. Chem. Phys.}}
	\textbf{\bibinfo{volume}{79}}, \bibinfo{pages}{1052} (\bibinfo{year}{1983}).
	
	\bibitem{Gast1983}
	\bibinfo{author}{Gast, A.~P.}, \bibinfo{author}{Hall, C.~K.} \&
	\bibinfo{author}{Russel, W.~B.}
	\newblock \bibinfo{title}{Polymer-induced phase separations in nonaqueous
		colloidal suspensions}.
	\newblock \emph{\bibinfo{journal}{J. Colloid Interface Sci.}}
	\textbf{\bibinfo{volume}{96}}, \bibinfo{pages}{251} (\bibinfo{year}{1983}).
	
	\bibitem{Sperry1984}
	\bibinfo{author}{Sperry, P.~R.}
	\newblock \bibinfo{title}{Morphology and mechanism in latex flocculated by
		volume restriction}.
	\newblock \emph{\bibinfo{journal}{J. Colloid Interface Sci.}}
	\textbf{\bibinfo{volume}{99}}, \bibinfo{pages}{97} (\bibinfo{year}{1984}).
	
	\bibitem{Lekkerkerker1992}
	\bibinfo{author}{Lekkerkerker, H. N.~W.}, \bibinfo{author}{Poon, W. C.~K.},
	\bibinfo{author}{Pusey, P.~N.}, \bibinfo{author}{Stroobants, A.} \&
	\bibinfo{author}{Warren, P.~B.}
	\newblock \bibinfo{title}{Phase behaviour of colloid + polymer mixtures}.
	\newblock \emph{\bibinfo{journal}{Europhys. Lett.}}
	\textbf{\bibinfo{volume}{20}}, \bibinfo{pages}{559--564}
	(\bibinfo{year}{1992}).
	
	\bibitem{Meijer1994}
	\bibinfo{author}{Meijer, E.~J.} \& \bibinfo{author}{Frenkel, D.}
	\newblock \bibinfo{title}{Colloids dispersed in polymer-solutions -- a
		computer-simulation study}.
	\newblock \emph{\bibinfo{journal}{J. Chem. Phys.}}
	\textbf{\bibinfo{volume}{100}}, \bibinfo{pages}{6873--6887}
	(\bibinfo{year}{1994}).
	
	\bibitem{Ilett1995}
	\bibinfo{author}{Ilett, S.~M.}, \bibinfo{author}{Orrock, A.},
	\bibinfo{author}{Poon, W. C.~K.} \& \bibinfo{author}{Pusey, P.~N.}
	\newblock \bibinfo{title}{Phase-behavior of a model colloid-polymer mixture}.
	\newblock \emph{\bibinfo{journal}{Phys. Rev. E}} \textbf{\bibinfo{volume}{51}},
	\bibinfo{pages}{1344--1352} (\bibinfo{year}{1995}).
	
	\bibitem{Dijkstra1999a}
	\bibinfo{author}{Dijkstra, M.}, \bibinfo{author}{Van~Roij, R.} \&
	\bibinfo{author}{Evans, R.}
	\newblock \bibinfo{title}{Phase diagram of highly asymmetric binary hard-sphere
		mixtures}.
	\newblock \emph{\bibinfo{journal}{Phys. Rev. E}} \textbf{\bibinfo{volume}{59}},
	\bibinfo{pages}{5744} (\bibinfo{year}{1999}).
	
	\bibitem{Bolhuis1994}
	\bibinfo{author}{Bolhuis, P.~G.} \& \bibinfo{author}{Frenkel, D.}
	\newblock \bibinfo{title}{Numerical study of the phase diagram of a mixture of
		spherical and rodlike colloids}.
	\newblock \emph{\bibinfo{journal}{J. Chem. Phys.}}
	\textbf{\bibinfo{volume}{101}}, \bibinfo{pages}{9869} (\bibinfo{year}{1994}).
	
	\bibitem{Petukhov2017}
	\bibinfo{author}{Petukhov, A.~V.}, \bibinfo{author}{Tuinier, R.} \&
	\bibinfo{author}{Vroege, G.~J.}
	\newblock \bibinfo{title}{Entropic patchiness: effects of colloid shape and
		depletion}.
	\newblock \emph{\bibinfo{journal}{Curr. Opin. Colloid Interface Sci.}}
	\textbf{\bibinfo{volume}{2}}, \bibinfo{pages}{600} (\bibinfo{year}{1997}).
	
	\bibitem{Glotzer2017}
	\bibinfo{author}{Du, C.~X.}, \bibinfo{author}{van Anders~G.},
	\bibinfo{author}{Newman, R.~S.} \& \bibinfo{author}{Glotzer, S.~C.}
	\newblock \bibinfo{title}{Shape-driven solid-solid transitions in colloids}.
	\newblock \emph{\bibinfo{journal}{PNAS}} \textbf{\bibinfo{volume}{114}},
	\bibinfo{pages}{E3892--E3899} (\bibinfo{year}{2017}).
	
	\bibitem{Lekkerkerker1994}
	\bibinfo{author}{Lekkerkerker, H. N.~W.} \& \bibinfo{author}{Stroobants, A.}
	\newblock \bibinfo{title}{Phase behaviour of rod-like colloid+ flexible polymer
		mixtures}.
	\newblock \emph{\bibinfo{journal}{Il Nuovo Cimento}}
	\textbf{\bibinfo{volume}{D16}}, \bibinfo{pages}{949} (\bibinfo{year}{1994}).
	
	\bibitem{Tuinier2007}
	\bibinfo{author}{Tuinier, R.}, \bibinfo{author}{Taniguchi, T.} \&
	\bibinfo{author}{Wensink, H.~H.}
	\newblock \bibinfo{title}{Phase behavior of a suspension of hard
		spherocylinders plus ideal polymer chains}.
	\newblock \emph{\bibinfo{journal}{Eur. Phys. J. E.}}
	\textbf{\bibinfo{volume}{23}}, \bibinfo{pages}{355} (\bibinfo{year}{2007}).
	
	\bibitem{Bolhuis1997a}
	\bibinfo{author}{Bolhuis, P.~G.}, \bibinfo{author}{Stroobants, A.},
	\bibinfo{author}{Frenkel, D.} \& \bibinfo{author}{Lekkerkerker, H. N.~W.}
	\newblock \bibinfo{title}{Numerical study of the phase behavior of rodlike
		colloids with attractive interactions}.
	\newblock \emph{\bibinfo{journal}{J. Chem. Phys.}}
	\textbf{\bibinfo{volume}{107}}, \bibinfo{pages}{1551} (\bibinfo{year}{1997}).
	
	\bibitem{Dogic1997}
	\bibinfo{author}{Dogic, Z.} \& \bibinfo{author}{Fraden, S.}
	\newblock \bibinfo{title}{Smectic phase in a colloidal suspension of
		semiflexible virus particles}.
	\newblock \emph{\bibinfo{journal}{Phys. Rev. Lett.}}
	\textbf{\bibinfo{volume}{78}}, \bibinfo{pages}{2417} (\bibinfo{year}{1997}).
	
	\bibitem{Edgar2002}
	\bibinfo{author}{Edgar, C.~D.} \& \bibinfo{author}{Gray, D.~G.}
	\newblock \bibinfo{title}{Influence of dextran on the phase behavior of
		suspensions of cellulose nanocrystals}.
	\newblock \emph{\bibinfo{journal}{Macromolecules}}
	\textbf{\bibinfo{volume}{35}}, \bibinfo{pages}{7400} (\bibinfo{year}{2002}).
	
	\bibitem{Buitenhuis1995}
	\bibinfo{author}{Buitenhuis, J.}, \bibinfo{author}{Donselaar, L.~N.},
	\bibinfo{author}{Buining, P.~A.}, \bibinfo{author}{Stroobants, A.} \&
	\bibinfo{author}{Lekkerkerker, H. N.~W.}
	\newblock \bibinfo{title}{Phase separation of mixtures of colloidal boehmite
		rods and flexible polymer}.
	\newblock \emph{\bibinfo{journal}{J. Colloid Interface Sci.}}
	\textbf{\bibinfo{volume}{175}}, \bibinfo{pages}{46} (\bibinfo{year}{1995}).
	
	\bibitem{Beck2007}
	\bibinfo{author}{Beck-Candanedo, S.}, \bibinfo{author}{Viet, D.} \&
	\bibinfo{author}{Gray, D.~G.}
	\newblock \bibinfo{title}{Triphase equilibria in cellulose nanocrystal
		suspensions containing neutral and charged macromolecules}.
	\newblock \emph{\bibinfo{journal}{Macromolecules}}
	\textbf{\bibinfo{volume}{40}}, \bibinfo{pages}{3429--3436}
	(\bibinfo{year}{2007}).
	
	\bibitem{Purdy2005}
	\bibinfo{author}{Purdy, K.~R.}, \bibinfo{author}{Varga, S.},
	\bibinfo{author}{Galindo, A.}, \bibinfo{author}{Jackson, G.} \&
	\bibinfo{author}{Fraden, S.}
	\newblock \bibinfo{title}{Nematic phase transitions in mixtures of thin and
		thick colloidal rods}.
	\newblock \emph{\bibinfo{journal}{Phys. Rev. Lett.}}
	\textbf{\bibinfo{volume}{94}}, \bibinfo{pages}{057801}
	(\bibinfo{year}{2005}).
	
	\bibitem{Kooij2000}
	\bibinfo{author}{Van~der Kooij, F.~M.}, \bibinfo{author}{Vogel, M.} \&
	\bibinfo{author}{Lekkerkerker, H. N.~W.}
	\newblock \bibinfo{title}{Phase behavior of a mixture of platelike colloids and
		nonadsorbing polymer}.
	\newblock \emph{\bibinfo{journal}{Phys. Rev. E}} \textbf{\bibinfo{volume}{62}},
	\bibinfo{pages}{5397} (\bibinfo{year}{2000}).
	
	\bibitem{Luan2009}
	\bibinfo{author}{Luan, L.}, \bibinfo{author}{Li, W.}, \bibinfo{author}{Liu, S.}
	\& \bibinfo{author}{Sun, D.}
	\newblock \bibinfo{title}{Phase behavior of mixtures of positively charged
		colloidal platelets and nonadsorbing polymer}.
	\newblock \emph{\bibinfo{journal}{Langmuir}} \textbf{\bibinfo{volume}{25}},
	\bibinfo{pages}{6349–6356} (\bibinfo{year}{2009}).
	
	\bibitem{Nakato2014}
	\bibinfo{author}{Nakato, T.}, \bibinfo{author}{Yamashita, Y.},
	\bibinfo{author}{Mouri, E.} \& \bibinfo{author}{Kuroda, K.}
	\newblock \bibinfo{title}{Multiphase coexistence and destabilization of liquid
		crystalline binary nanosheet colloids of titanate and clay}.
	\newblock \emph{\bibinfo{journal}{Soft Matter}} \textbf{\bibinfo{volume}{10}},
	\bibinfo{pages}{3161} (\bibinfo{year}{2014}).
	
	\bibitem{Chen2015}
	\bibinfo{author}{Chen, M.} \emph{et~al.}
	\newblock \bibinfo{title}{Observation of isotropic-–isotropic demixing in
		colloidal platelet-sphere mixtures}.
	\newblock \emph{\bibinfo{journal}{Soft Matter}} \textbf{\bibinfo{volume}{11}},
	\bibinfo{pages}{5775} (\bibinfo{year}{2015}).
	
	\bibitem{Woolston2015}
	\bibinfo{author}{Woolston, P.} \& \bibinfo{author}{van Duijneveldt, J.~S.}
	\newblock \bibinfo{title}{Three-phase coexistence in colloidal rod-plate
		mixtures}.
	\newblock \emph{\bibinfo{journal}{Langmuir}} \textbf{\bibinfo{volume}{31}},
	\bibinfo{pages}{9290--9295} (\bibinfo{year}{2015}).
	
	\bibitem{Lekkerkerker1990}
	\bibinfo{author}{Lekkerkerker, H. N.~W.}
	\newblock \bibinfo{title}{Osmotic equilibrium treatment of the phase separation
		in colloidal dispersions containing non-adsorbing polymer molecules}.
	\newblock \emph{\bibinfo{journal}{Colloids Surf.}}
	\textbf{\bibinfo{volume}{51}}, \bibinfo{pages}{419} (\bibinfo{year}{1990}).
	
	\bibitem{Aarts2002}
	\bibinfo{author}{Aarts, D. G. A.~L.}, \bibinfo{author}{Tuinier, R.} \&
	\bibinfo{author}{Lekkerkerker, H. N.~W.}
	\newblock \bibinfo{title}{Phase behaviour of mixtures of colloidal spheres and
		excluded-volume polymer chains}.
	\newblock \emph{\bibinfo{journal}{J. Phys: Condens. Matter}}
	\textbf{\bibinfo{volume}{14}}, \bibinfo{pages}{7551} (\bibinfo{year}{2002}).
	
	\bibitem{Fleer2008}
	\bibinfo{author}{Fleer, G.~J.} \& \bibinfo{author}{Tuinier, R.}
	\newblock \bibinfo{title}{Analytical phase diagrams for colloids and
		non-adsorbing polymer}.
	\newblock \emph{\bibinfo{journal}{Adv. Colloid Interface Sci.}}
	\textbf{\bibinfo{volume}{143}}, \bibinfo{pages}{1--47}
	(\bibinfo{year}{2008}).
	
	\bibitem{Vrij1976}
	\bibinfo{author}{Vrij, A.}
	\newblock \bibinfo{title}{Polymers at interfaces and the interactions in
		colloidal dispersions}.
	\newblock \emph{\bibinfo{journal}{Pure Appl. Chem.}}
	\textbf{\bibinfo{volume}{48}}, \bibinfo{pages}{471--483}
	(\bibinfo{year}{1976}).
	
	\bibitem{Eisenriegler1997}
	\bibinfo{author}{Eisenriegler, E.}
	\newblock \bibinfo{title}{Universal density-force relations for polymers near a
		repulsive wall}.
	\newblock \emph{\bibinfo{journal}{Phys. Rev. E}} \textbf{\bibinfo{volume}{55}},
	\bibinfo{pages}{3116} (\bibinfo{year}{1997}).
	
	\bibitem{Tuinier2000}
	\bibinfo{author}{Tuinier, R.}, \bibinfo{author}{Vliegenthart, G.~A.} \&
	\bibinfo{author}{Lekkerkerker, H. N.~W.}
	\newblock \bibinfo{title}{Depletion interaction between spheres immersed in a
		solution of ideal polymer chains}.
	\newblock \emph{\bibinfo{journal}{J. Chem. Phys.}}
	\textbf{\bibinfo{volume}{113}}, \bibinfo{pages}{10768--10775}
	(\bibinfo{year}{2000}).
	
	
	
	\bibitem{Bates2000}
	\bibinfo{author}{Bates, M.~A.} \& \bibinfo{author}{Frenkel, D.}
	\newblock \bibinfo{title}{Phase behavior of model mixtures of colloidal disks and polymers}.
	\newblock \emph{\bibinfo{journal}{Phys.  Rev.  E}} \textbf{\bibinfo{volume}{62}},
	\bibinfo{pages}{5225-5229} (\bibinfo{year}{2000}).
	
	
	\bibitem{Zhang2002}
	\bibinfo{author}{Zhang, S.~D.}, \bibinfo{author}{Reynolds, P.~A.} \&
	\bibinfo{author}{van Duijneveldt, J.~S.}
	\newblock \bibinfo{title}{Phase behavior of mixtures of colloidal platelets and
		nonadsorbing polymers}.
	\newblock \emph{\bibinfo{journal}{J. Chem. Phys.}}
	\textbf{\bibinfo{volume}{117}}, \bibinfo{pages}{9947} (\bibinfo{year}{2002}).
	
	\bibitem{Zhang2002a}
	\bibinfo{author}{Zhang, S.~D.}, \bibinfo{author}{Reynolds, P.~A.} \&
	\bibinfo{author}{van Duijneveldt, J.~S.}
	\newblock \bibinfo{title}{Phase separation in mixtures of colloidal platelets
		and non-adsorbing polymer: a scaled particle treatment}.
	\newblock \emph{\bibinfo{journal}{Mol. Phys.}} \textbf{\bibinfo{volume}{100}},
	\bibinfo{pages}{3041} (\bibinfo{year}{2002}).
	
	\bibitem{Aliabadi2016}
	\bibinfo{author}{Aliabadi, R.}, \bibinfo{author}{Moradi, M.} \&
	\bibinfo{author}{Varga, S.}
	\newblock \bibinfo{title}{Tracking three-phase coexistences in binary mixtures
		of hard plates and spheres}.
	\newblock \emph{\bibinfo{journal}{J. Chem. Phys.}}
	\textbf{\bibinfo{volume}{144}}, \bibinfo{pages}{074902}
	(\bibinfo{year}{2016}).
	
	\bibitem{Wensink2004}
	\bibinfo{author}{Wensink, H.~H.} \& \bibinfo{author}{Lekkerkerker, H. N.~W.}
	\newblock \bibinfo{title}{Sedimentation and multi-phase equilibria in mixtures
		of platelets and ideal polymer}.
	\newblock \emph{\bibinfo{journal}{Europhys. Lett.}}
	\textbf{\bibinfo{volume}{66}}, \bibinfo{pages}{125--131}
	(\bibinfo{year}{2004}).
	
	\bibitem{Cox2001}
	\bibinfo{author}{Cox, D.~E.} \emph{et~al.}
	\newblock \bibinfo{title}{Universal phase diagram for high-piezoelectric
		perovskite systems}.
	\newblock \emph{\bibinfo{journal}{Appl. Phys. Lett.}}
	\textbf{\bibinfo{volume}{79}}, \bibinfo{pages}{400--402}
	(\bibinfo{year}{2001}).
	
	\bibitem{Ahart2008}
	\bibinfo{author}{Ahart, M.} \emph{et~al.}
	\newblock \bibinfo{title}{Origin of morphotropic phase boundaries in
		ferroelectrics}.
	\newblock \emph{\bibinfo{journal}{Nature}} \textbf{\bibinfo{volume}{451}},
	\bibinfo{pages}{545--548} (\bibinfo{year}{2008}).
	
	\bibitem{Park2013}
	\bibinfo{author}{Park, J.~H.} \emph{et~al.}
	\newblock \bibinfo{title}{Measurement of a solid-state triple point at the
		metal-insulator transition in $\text{VO}$$_2$}.
	\newblock \emph{\bibinfo{journal}{Nature}} \textbf{\bibinfo{volume}{500}},
	\bibinfo{pages}{431--434} (\bibinfo{year}{2013}).
	
	\bibitem{Park2010}
	\bibinfo{author}{Park, K.}, \bibinfo{author}{Koerner, H.} \&
	\bibinfo{author}{Vaia, R.~A.}
	\newblock \bibinfo{title}{Depletion-induced shape and size selection of gold
		nanoparticles}.
	\newblock \emph{\bibinfo{journal}{Nano Lett.}} \textbf{\bibinfo{volume}{10}},
	\bibinfo{pages}{1433--1439} (\bibinfo{year}{2010}).
	
	\bibitem{Saunders2008}
	\bibinfo{author}{Saunders, B.~R.} \& \bibinfo{author}{Turner, M.~L.}
	\newblock \bibinfo{title}{Nanoparticle-–polymer photovoltaic cells}.
	\newblock \emph{\bibinfo{journal}{Adv. Colloid Interface Sci.}}
	\textbf{\bibinfo{volume}{138}}, \bibinfo{pages}{1--23}
	(\bibinfo{year}{2008}).
	
	\bibitem{Kouijzer2013}
	\bibinfo{author}{Kouijzer, S.} \emph{et~al.}
	\newblock \bibinfo{title}{Predicting morphologies of solution processed polymer
		: Fullerene blends}.
	\newblock \emph{\bibinfo{journal}{J. Am. Chem. Soc}}
	\textbf{\bibinfo{volume}{135}}, \bibinfo{pages}{12057--12067}
	(\bibinfo{year}{2013}).
	
	\bibitem{Wu2007}
	\bibinfo{author}{Wu, J.} \& \bibinfo{author}{Li, Z.}
	\newblock \bibinfo{title}{Density--functional theory for complex fluids}.
	\newblock \emph{\bibinfo{journal}{Annu. Rev. Phys. Chem.}}
	\textbf{\bibinfo{volume}{58}}, \bibinfo{pages}{85--112}
	(\bibinfo{year}{2007}).
	
	\bibitem{Parsons1979}
	\bibinfo{author}{Parsons, J.~D.}
	\newblock \bibinfo{title}{Nematic ordering in a system of rods}.
	\newblock \emph{\bibinfo{journal}{Phys. Rev. A}} \textbf{\bibinfo{volume}{19}},
	\bibinfo{pages}{1225--1230} (\bibinfo{year}{1979}).
	
	\bibitem{Lee1987}
	\bibinfo{author}{Lee, S.~D.}
	\newblock \bibinfo{title}{A numerical investigation of nematic ordering based
		on a simple hard rod model}.
	\newblock \emph{\bibinfo{journal}{J. Chem. Phys.}}
	\textbf{\bibinfo{volume}{87}}, \bibinfo{pages}{4972--4974}
	(\bibinfo{year}{1987}).
	
	\bibitem{Odijk1986}
	\bibinfo{author}{Odijk, T.}
	\newblock \bibinfo{title}{Theory of lyotropic polymer liquid crystals}.
	\newblock \emph{\bibinfo{journal}{Macromolecules}}
	\textbf{\bibinfo{volume}{19}}, \bibinfo{pages}{2313} (\bibinfo{year}{1986}).
	
	\bibitem{Vroege1992}
	\bibinfo{author}{Vroege, G.~J.} \& \bibinfo{author}{Lekkerkerker, H. N.~W.}
	\newblock \bibinfo{title}{Phase transitions in lyotropic colloidal and polymer
		liquid crystals}.
	\newblock \emph{\bibinfo{journal}{Rep. Progr. Phys.}}
	\textbf{\bibinfo{volume}{55}}, \bibinfo{pages}{1241} (\bibinfo{year}{1992}).
	
	\bibitem{Lennard1937}
	\bibinfo{author}{Lennard-Jones, J.~E.} \& \bibinfo{author}{Devonshire, A.~F.}
	\newblock \bibinfo{title}{Critical phenomena in gases}.
	\newblock \emph{\bibinfo{journal}{Proc. Roy. Soc.}}
	\textbf{\bibinfo{volume}{163A}}, \bibinfo{pages}{53} (\bibinfo{year}{1937}).
	
	\bibitem{Malijevski}
	\bibinfo{author}{Malijevsk\'{y}, A.} \& \bibinfo{author}{Kolafa, J.}
	\newblock \emph{\bibinfo{title}{Introduction to the Thermodynamics of Hard
			Spheres and Related Systems, in: 'Theory and Simulation of Hard-Sphere Fluids
			and Related Systems', Lecture Notes in Physics, vol. 753}}
	(\bibinfo{publisher}{Springer}, \bibinfo{address}{Berlin, Heidelberg},
	\bibinfo{year}{2008}).
	
\end{thebibliography}


\end{document}